\documentclass[prl,twocolumn,showpacs,superscriptaddress,floatfix]{revtex4}

\usepackage{graphicx}
\usepackage{amssymb}
\usepackage{amsbsy}

\begin{document}

\title{A Corrsin type approximation for Lagrangian fluid Turbulence}
\author{Rudolf \surname{Friedrich}}
\affiliation{Theoretische Physik, Universit\"at M\"unster, Germany}
\author{Rainer \surname{Grauer}}
\affiliation{Theoretische Physik I, Ruhr-Universit\"at Bochum, Germany}
\author{Holger \surname{Homann}}
\affiliation{Theoretische Physik I, Ruhr-Universit\"at Bochum, Germany}
\author{Oliver \surname{Kamps}}
\affiliation{Theoretische Physik, Universit\"at M\"unster, Germany}
%\email{grauer@tp1.rub.de}

\date{\today}

\begin{abstract}
  In Lagrangian turbulence one is faced with the puzzle that 2D
  Navier-Stokes flows are nearly as intermittent as in three
  dimensions although no intermittency is present in the inverse
  cascade in 2D Eulerian turbulence. In addition, an inertial range is
  very difficult to detect and it is questionable whether it exists at
  all. Here, we investigate the transition of Eulerian to Lagrangian
  probability density functions (PDFs) which leads to a new type of
  Lagrangian structure function. This possesses an extended inertial
  range similar to the case of tracer particles in a frozen turbulent
  velocity field. This allows a connection to the scaling of Eulerian
  transversal structure functions.
\end{abstract}


% 47.10.ad Navier-Stokes equations
% 47.27.-i Turbulent flows
% 47.27.E- Turbulence simulation and modeling
% 02.50.Fz Stochastic analysis

\maketitle

\noindent
\textit{Introduction} The Lagrangian treatment of fluids in turbulence
has gained tremendous attraction in the last 10 years. Prominent
examples are the passive scalar advection and statistical conservation
laws (see e.g. Falkovich and Sreenivasan
\cite{falkovich-sreenivasan:2006} and references therein) and the
dynamics of tracer particles in incompressible flows.  Especially, the
latter has undergone a rapid development due to the enormous progress
in experimental techniques measuring particle trajectories
\cite{ott-mann:2000, porta-voth-etal:2000, porta-voth-etal:2001a,
  voth-porta-etal:2001b, mordant-metz-etal:2001,
  mordant-leveque-etal:2004}.  Motivated by these experiments,
numerical simulations \cite{biferale-bofetta-etal:2004,
  kamps-friedrich:2007, homann-grauer-etal:2007}, multifractal and
PDF-modeling \cite{borgas:1993, biferale-bofetta-etal:2004,
  chevillard-roux-etal:2003, xu-quellette-etal:2006, friedrich:2003}
have revealed many interesting features of Lagrangian turbulence as
e.g. the distinguished role of coherent structures or nearly singular
structures compared to the Eulerian description. Despite this
progress, simple and basic questions in Lagrangian turbulence remain
open. In this Letter, we first state some of these open problems. We
show that applying a type of Corrsin approximation \cite{corrsin:1959}
in order to relate Eulerian two point velocity increments and single
particle Lagrangian velocity increments the definition of a new type
of Lagrangian structure function becomes necessary.  The properties of
this new type of structure functions are first studied on the simpler
case of advection of tracer particles in a frozen velocity field.
Here, it is possible to relate the Lagrangian structure functions to
the transverse Eulerian ones. Finally, the full dynamical new
structure functions are investigated using high resolution spectral
simulations ($1024^3$ mesh points in 3D, $1024^2$ mesh points in 2D)
and their scaling behavior is discussed.

\begin{figure}
  \centering
  \includegraphics[width=.49\textwidth]{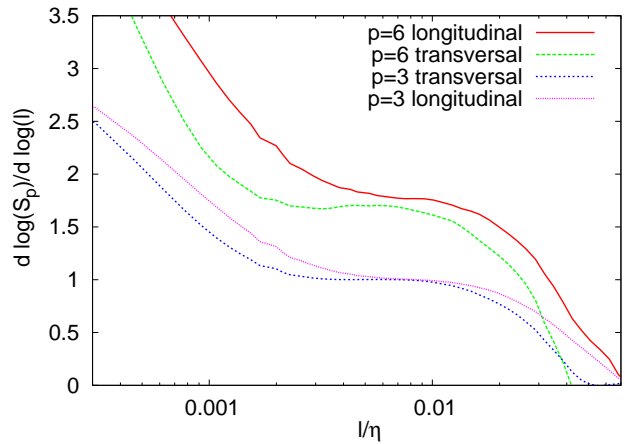}\\
  \caption{Logarithmic derivative of 3D Eulerian longitudinal and
    transverse structure functions}
  \label{euler_long_trans}
\end{figure}

\begin{figure}
  \centering
  \includegraphics[width=.49\textwidth]{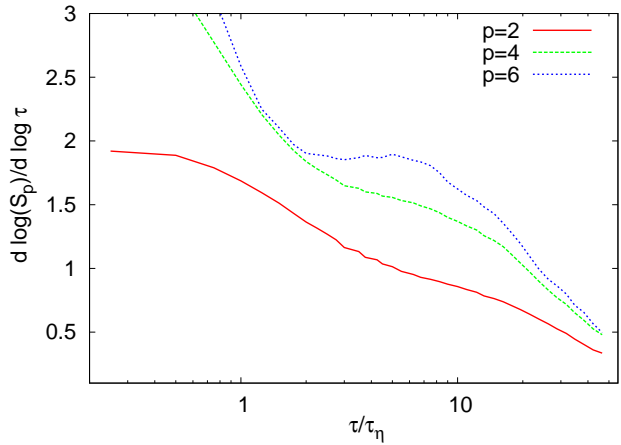}
  \caption{Logarithmic derivative of 3D Lagrangian structure
    functions}
  \label{badscaling}
\end{figure}

\begin{figure}
  \centering
  \includegraphics[width=.49\textwidth]{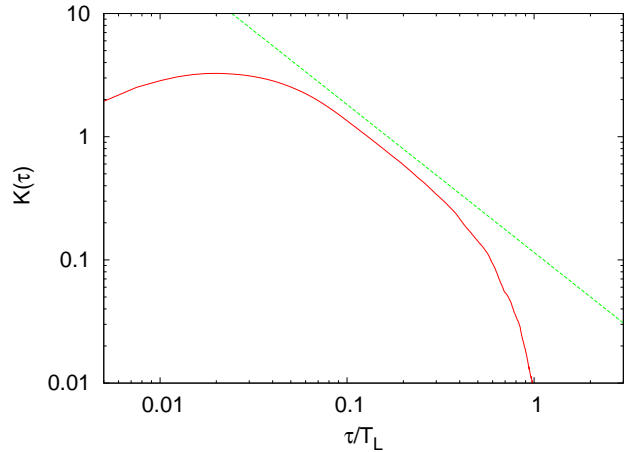}
  \caption{Excess curtosis $K(\tau)$ of 2D Lagrangian turbulence}
\label{curtosis}
\end{figure}

\noindent
\textit{Description of open problems} The basic open problems in
Lagrangian turbulence that we want to address are: i)~a dramatically
decreased scaling range for the Lagrangian structure functions
compared to their Eulerian counterparts and ii)~missing monotonicity
in relating Eulerian to Lagrangian turbulence. In the following, we
describe in more detail the meaning of these points.

\noindent
i) One major problem in using the usual Lagrangian velocity increments
$\mathbf{u}(\mathbf{x}(\mathbf{y},t),t) - \mathbf{u}(\mathbf{y},0)$ is
their dramatically decreased scaling range compared to their Eulerian
counterparts. This issue is visible in Fig.~\ref{euler_long_trans} and
Fig.~\ref{badscaling}. Here, the logarithmic derivatives for Eulerian
and Lagrangian structure functions are shown. For the latter it is
hard to detect a clear scaling range. Using extended self-similarity
(ESS \cite{benzi-ciliberto-etal:1993}) the situation gets slightly
better. Biferale \textit{et al.}  \cite{biferale-bofetta-etal:2004}
have chosen the interval $10 \tau_\eta \leq \tau \leq 50 \tau_\eta$ as
a possible scaling range whereas in the experiments
\cite{porta-voth-etal:2001a}, \cite{mordant-leveque-etal:2004} and the
simulation \cite{homann-grauer-etal:2007} smaller temporal increments
have been chosen (approximately $3 \tau_\eta \leq \tau \leq 15
\tau_\eta$) where $\tau_\eta$ denotes the Kolmogorov dissipation time.
Clearly, in each of this different temporal regions, different physics
takes place: for large time lags, a tendency to Gaussian behavior sets
in whereas strong vortices influence the intermittency for smaller
time increments. Therefore, the question arises where to locate the
inertial range and even the question whether there is any scaling
range at all must be taken seriously.

\noindent
ii) The present attempts to describe Lagrangian intermittency mostly
translate features from the Eulerian viewpoint in a simple way to the
Lagrangian one. The first step is an assumption how to relate velocity
increments in time $\delta_\tau v$ to velocity increments in space
$\delta_r u$. The assumption used in \cite{biferale-bofetta-etal:2004}
is a Kolmogorov type argument that $r$ and $\tau$ are related by the
typical eddy turnover time on that scale: $\tau \sim r/\delta_r u$.
With this identification, the Eulerian multifractal modeling can be
translated to the Lagrangian one yielding predictions for Lagrangian
structure functions and PDFs of the acceleration. The important
ingredient is the Eulerian singularity spectrum
(\cite{chevillard-roux-etal:2003}, \cite{biferale-bofetta-etal:2004}).
A direct consequence of this type of modeling is the observation that
if one considers two different turbulent systems (e.g.
magnetohydrodynamic versus Navier-Stokes turbulence or 3D versus 2D
Navier-Stokes turbulence) where one system is more intermittent than
the other in the Eulerian framework, than this system should also be
more intermittent in the Lagrangian framework.  We call this feature
the monotonicity property. For the case of MHD turbulence this has
been shown not to be true \cite{homann-grauer-etal:2007}. The problem
is even more visible if one compares scaling in the inverse energy
cascade in Navier-Stokes flows in two dimensions with the direct
cascade in three dimensions.  It turns out (see Kamps and Friedrich
\cite{kamps-friedrich:2007}) that although the inverse cascade shows
no intermittency in the Eulerian framework, the Lagrangian
intermittency is as strong as in the three-dimensional case. In order
to characterize intermittency we have determined the excess curtosis
of the Lagrangian velocity increment time difference (see Fig.
\ref{curtosis}). The Lagrangian curtosis is definitely different from
a non-intermittent situation exhibiting clear scaling behavior. It
decays to zero for large time differences with a powerlaw with scaling
exponent of about 1.2~.  In addition, calculating Lagrangian structure
functions in the same manner as in \cite{homann-grauer-etal:2007}
yields similar values for the scaling exponents (see
\cite{kamps-friedrich:2007}).

\noindent

\noindent
\textit{New Lagrangian structure function} 
A first attempt to relate Eulerian and Lagrangian statistics has been
made by Corrsin \cite{corrsin:1959}. Although the Corrsin
approximation has serious shortcomings, as has been recently discussed
by Ott and Mann \cite{ott-mann:2005}, it can serve as a first
guideline. Let us consider the PDF $ f({\bf v},t)$ of the Lagrangian
velocity increment
\begin{equation}
\label{standardInc}
{\bf v}({\bf y},t)={\bf u}({\bf x}({\bf y},t),t) - {\bf u}({\bf y},0) \;,
\end{equation}
where ${\bf x}({\bf y},t)=\tilde{\bf x}({\bf y},t)+{\bf y}$ denotes
the particle path starting at position ${\bf y}$ at time $t=0$, i.e.
$\tilde {\bf x}({\bf y},t=0)=0$:
\begin{eqnarray}
  f({\bf v},t) &=& \langle \delta({\bf v}-
  [{\bf u}({\bf x}({\bf y},t),t)-
  {\bf u}({\bf y},0)]) \rangle \\
  &=&\int d{\bf x} \langle\delta({\bf x}-{\bf x}({\bf y},t))
  \delta({\bf v}-
  [{\bf u}({\bf x},t)-{\bf u}({\bf y},0)])\rangle \; .
  \nonumber
\end{eqnarray}
We assume homogeneous turbulence such that the dependence on the
starting position ${\bf y}$ vanishes. The Corrsin approximation
relies on the assumption of statistical independency of the path ${\bf
  x}({\bf y},t)$ and the velocity difference ${\bf u}({\bf x},t)-{\bf
  u}({\bf y},0)$ leading to the factorization of the expectation
value:
\begin{equation}\label{Corrsin}
  f({\bf v},t)=\int d{\bf x} \, p({\bf x};{\bf y},t) 
  \langle\delta({\bf v}-
  [{\bf u}({\bf x},t)-{\bf u}({\bf y},0)]) \rangle
\end{equation}
Here, the transition probability $p({\bf x};{\bf y},t)$,
\begin{equation}
  p({\bf x};{\bf y},t)=\langle\delta({\bf x}-{\bf x}({\bf y},t))\rangle
\end{equation}
has been introduced. In this approximation, no direct relationship between
the PDF of the Lagrangian and Eulerian velocity increments can be 
established. However, formula (\ref{Corrsin}) suggests to define the novel 
velocity increment
\begin{equation}
  \label{novelInc}
  {\bf w}({\bf y},t)={\bf u}({\bf y}+\tilde{\bf x}({\bf y},t),t)-
  {\bf u}({\bf y},t)
\end{equation}
In this case a direct connection of Eulerian and Lagrangian velocity
increments becomes possible:
\begin{equation}\label{exact} 
  f({\bf w},t)=\int d{\bf x} \, p({\bf x},{\bf y}|{\bf w},t) 
  \langle\delta({\bf w}-[{\bf u}({\bf x},t)-{\bf u}({\bf y},t)])\rangle
\end{equation}
Here, we have introduced the conditional transition
probability for a particle starting at t=0 at position {\bf y} and
ending at time t at position {\bf x} under the condition that the
velocity increment across scale ${\bf x}-{\bf y}$ is just ${\bf w}$.
Corrsin's approximation to the exact expression (\ref{exact}) replaces
this conditional probability distribution by the unconditioned one.
We want to point out that a similar correspondence between Eulerian
velocity increment ${\bf v}_E({\bf r},t)$ and Lagrangian velocity
increment ${\bf v}({\bf y},t)$ can not be established in such a
straightforward manner.
A further property, which is relevant for two dimensional fluid turbulence,
follows directly from the exact relation (\ref{exact}), which relates 
Lagrangian and Eulerian increment statistics. If the Eulerian statistics
does not exhibit intermittency whereas Lagrangian statistics does,
the signatures of intermittency are contained in the conditional
probability distribution $p({\bf x},{\bf y}|{\bf w},t)$.

The definition of the Lagrangian velocity increment (\ref{novelInc})
actually focuses on the fluctuations related to the motion of
the particle in the turbulent field. In contrast to the usual
definition, there is no contribution from the fluctuations due to
the temporal evolution of the velocity field at two different 
spatial points. Therefore, one expects that the statistics of this
increment is close to the one obtained from the motion of a
Lagrangian particle in a frozen turbulent field. For two dimensional
turbulence, thereby, the motion of the Lagrangian particles are integrable
and related to the lines of constant stream function.  

\begin{figure}
  \begin{center}
    \includegraphics[width=.49\textwidth]{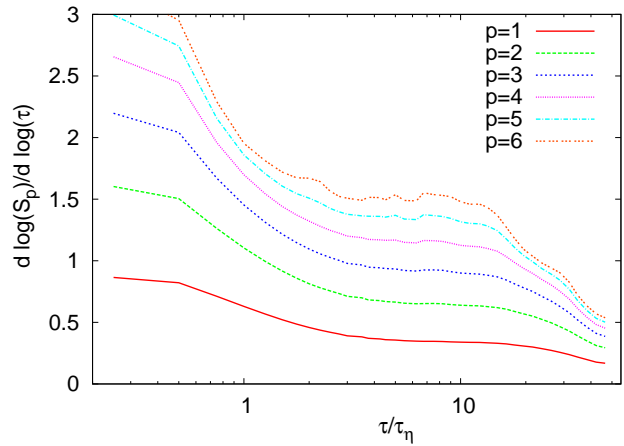}
    \caption{Logarithmic derivative of Lagrangian structure functions
    in frozen 3D Navier-Stokes turbulence, $R_\lambda=316$}
    \label{frozen_diffLog}
  \end{center}
\end{figure}

\begin{figure}
  \begin{center}
    \includegraphics[width=.49\textwidth]{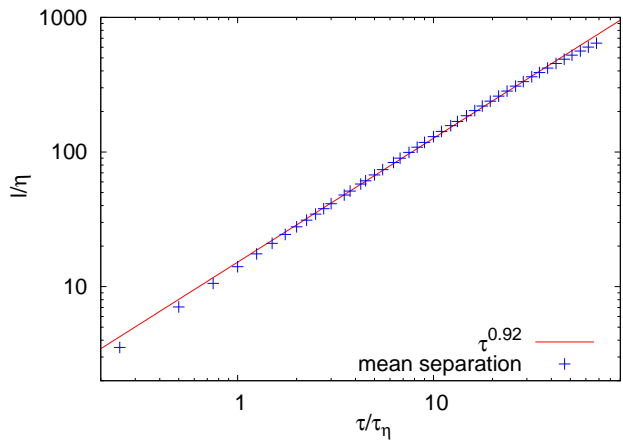}
    \caption{Mean separation $l$ as function of the time increment~$\tau$ }
    \label{ltotau}
  \end{center}
\end{figure}

\begin{figure}
  \begin{center}
    \includegraphics[width=.4\textwidth]{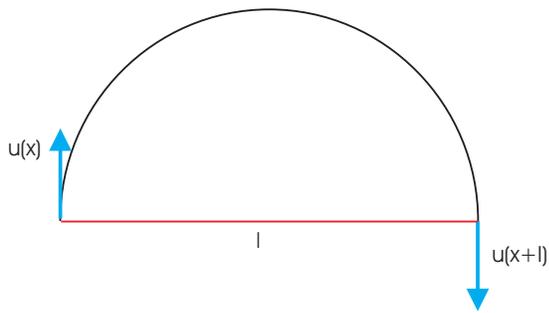}
    \caption{Sketch of the velocity increment over a half circle like
    trajectory}
    \label{vortex}
  \end{center}
\end{figure}

\noindent
\textit{Case I: Frozen turbulence}
The proposed novel velocity increment naturally occurs in frozen
turbulence. Here, the tracer particles are advanced in a static
velocity field. In this scenario the novel velocity increment
(\ref{novelInc}) and the standard Lagrangian velocity increment
(\ref{standardInc}) are identical. The measured structure functions of
several orders are given in Fig.~\ref{frozen_diffLog}. They exhibit an
extended inertial range, whose size is approximately as large as in
Eulerian turbulence, if the time-lag is transformed into spatial
differences. The scaling exponents of frozen turbulence are given in
Table.~\ref{table1}. They can be explained as follows: The main
contributions to the structure functions originate from half circles
like trajectories. The starting and ending points of these are
separated by a distance $l$ (see Fig.~\ref{vortex}). In this
configuration the velocity increment (\ref{novelInc}) corresponds to
the transverse Eulerian velocity increment over this distance $l$. The
corresponding structure functions are shown in
Fig.~\ref{euler_long_trans}. The transverse statistics are slightly
more intermittent, as was already observed by
\cite{gotoh-fukayama-etal:2002}. It is still under discussion if this
might be a finite Reynolds number effect. However, the Lagrangian
scaling exponents $\zeta_p^F $ of frozen turbulence and the Eulerian
transverse exponents $\zeta_p^T$ fulfill the relation $\zeta_p^F =
(\zeta_p^T)^{0.92}$, as can be seen in Table~\ref{table1}. The
exponent $0.92$ comes from the scaling law of the mean separation $l$
within the inertial range. As is shown in Fig.~\ref{ltotau} $l$ scales
as $\tau^{0.92}$.
\begingroup
\squeezetable
\begin{table}[!h]
  \centering
  \begin{ruledtabular}
  \begin{tabular}{cccc}
    order p & $\zeta_p^F$     &  $\zeta_p^T\cdot 0.92$ & $\zeta_p^N$ \\\hline
    1       & 0.35 $\pm$ 0.03 & 0.34 $\pm$ 0.02        & 0.35 $\pm$ 0.03\\
    2       & 0.66 $\pm$ 0.03 & 0.64 $\pm$ 0.05        & 0.65 $\pm$ 0.04\\
    3       & 0.92 $\pm$ 0.04 & 0.91 $\pm$ 0.07        & 0.91 $\pm$ 0.03\\
    4       & 1.15 $\pm$ 0.05 & 1.14 $\pm$ 0.11        & 1.14 $\pm$ 0.02\\
    5       & 1.34 $\pm$ 0.06 & 1.35 $\pm$ 0.13        & 1.34 $\pm$ 0.01\\
    6       & 1.49 $\pm$ 0.07 & 1.52 $\pm$ 0.15        & 1.50 $\pm$ 0.02\\
    7       & 1.63 $\pm$ 0.08 & 1.67 $\pm$ 0.18        & 1.63 $\pm$ 0.05\\
    8       & 1.73 $\pm$ 0.07 & 1.78 $\pm$ 0.21        & 1.73 $\pm$ 0.08
  \end{tabular}
  \end{ruledtabular}
  \caption{Scaling exponents: Frozen Lagrangian $\zeta_p^F$; Eulerian
    transversal $\zeta_p^T$; Novel Lagrangian $\zeta_p^N$ }
  \label{table1}
\end{table}
\endgroup
\begingroup
\squeezetable
\begin{table}[!h]
  \centering
  \begin{ruledtabular}
  \begin{tabular}{ccccccc}
     order p &  $\zeta_p^N$     &  $\zeta_p^F$ \\\hline
      1      & 0.53 $\pm$ 0.01  &  0.52 $\pm$ 0.01 \\
      2      & 1                & 1 \\
      3      & 1.44 $\pm$ 0.01  & 1.45 $\pm$ 0.01 \\
      4      & 1.81 $\pm$ 0.05  & 1.83 $\pm$ 0.06 \\
      5      & 2.06 $\pm$ 0.13  & 2.09 $\pm$ 0.14 \\
      6      & 2.18 $\pm$ 0.22  & 2.23 $\pm$ 0.23 \\
  \end{tabular}
  \end{ruledtabular}
  \caption{ESS Scaling exponents for 2D Turbulence: Novel Lagrangian $\zeta_p^N$; Frozen Lagrangian $\zeta_p^F$ }
  \label{table2}
\end{table}
\endgroup
\\
\noindent
\textit{Case II: Dynamical turbulence}
We have determined the scaling exponents of the new Lagrangian
structure function for two and three dimensions.  For the two
dimensional case scaling exponents have been extracted by the method of
ESS.  Our observation is the existence of a wider scaling range as
compared to the case of the standard Lagrangian increments.
Furthermore, there is close agreement of the scaling exponents with the
case of frozen turbulent fields.  Within the errorbars the values in
Table.~\ref{table2} show an agreement between the new increments and
the standard Lagrangian increments for the frozen field.  Furthermore,
we have determined the corresponding PDFs. Whereas there is a
discrepancy between the PDF obtained from frozen turbulence and the
standard Lagrangian velocity, the PDF of the
new increment coincides with the one obtained from frozen turbulence.

\begin{figure}
  \centering
  \includegraphics[width=.49\textwidth]{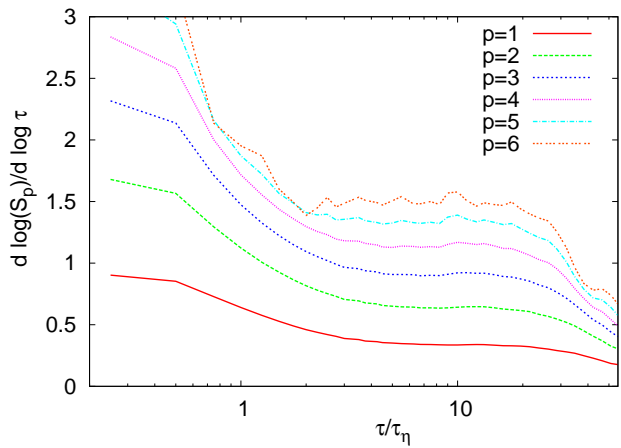}
  \caption{Logarithmic derivative of the new Lagrangian structure
    functions for 3D Navier-Stokes turbulence, $R_\lambda =316$}
  \label{LETIdiffLog}
\end{figure}

\noindent
{\it Conclusions and outlook}
An important issue in turbulence research is to establish relations
between Eulerian and Lagrangian statistics.  In the present Letter, we
have examined possible relations between Eulerian and Lagrangian
intermittency. 
% Our numerical calculations have led us to critizise
% arguments based on the multifractal model.
Furthermore, we have defined a new Lagrangian velocity increment,
which directly focuses on the fluctuations picked up by a moving
Lagrangian particle. This new increment exhibits a more pronounced
scaling regime. Its scaling behavior is equal to the one obtained for
frozen turbulence and can be related to the Eulerian transverse
structure functions.  In fact, the probability distributions of this
new type of increment and the one obtained from frozen turbulence
coincide. This has been verified for three dimensional as well as two
dimensional turbulence.

\noindent
{\it Acknowledgments.---}
Access to the JUMP multiprocessor computer at the FZ J\"ulich was made
available through project HB022. Part of the computations were
performed on an Linux-Opteron cluster supported by HBFG-108-291. The
work of H.H. and R.G. benefited from support through SFB 591 of the
Deutsche Forschungsgesellschaft. R.F. and O. K. acknowledge
support from the Deutsche Forschungsgesellschaft (FR 1003/8-1).


\begin{thebibliography}{99}
%\bibliographystyle{apsrev}
\bibitem{falkovich-sreenivasan:2006}
G. Falkovich and K.R. Sreenivasan,
Physics Today \textbf{59}, 43 (2006)

\bibitem{ott-mann:2000}
S.~Ott and J. Mann,
J. Fluid Mech. \textbf{422}, 207 (2000).

\bibitem{porta-voth-etal:2000}
A. La Porta, G. Voth, F. Moisy, and E. Bodenschatz,
Phys. Fluids \textbf{12}, 1485 (2000).

\bibitem{porta-voth-etal:2001a}
A. La Porta, G. Voth, A.M. Crawford, J. Alexander, and E. Bodenschatz,
Nature \textbf{409}, 1017 (2001).

\bibitem{voth-porta-etal:2001b}
G. Voth, A. La Porta, A.M. Crawford, J. Alexander, and E. Bodenschatz,
Rev. Sci. Instr. \textbf{12}, 4348 (2001).

\bibitem{mordant-metz-etal:2001}
N. Mordant, P. Metz, O. Michel, and J.-F. Pinton,
Phys. Rev. Lett \textbf{87}, 214501 (2001).

\bibitem{mordant-leveque-etal:2004}
N. Mordant, E. L\'ev\^eque, and J. F. Pinton,
New Journ. Phys., \textbf{6} 116 (2004).

\bibitem{biferale-bofetta-etal:2004}
L.~Biferale, G.~Bofetta, A.~Celani, B.J.~Devinish, A. Lanotte, and F. Toschi,
Phys. Rev. Lett. \textbf{93}, 064502 (2004).

\bibitem{kamps-friedrich:2007} O. Kamps and R. Friedrich, in preparation (2007).

\bibitem{homann-grauer-etal:2007}
H. Homann, R. Grauer, A. Busse, and W.C. M\"uller,
to appear in J. Plasma Phys, arXiv:physics/0702115.

\bibitem{borgas:1993}
M.S. Borgas,
Philos. Trans. R. Soc. Lond. A \textbf{342} 379 (1993).

\bibitem{chevillard-roux-etal:2003}
L. Chevillard, S. G. Roux, E. L\'ev\^eque, N. Mordant, J.-F. 
Pinton, and A. Arn\'eodo,
Phys. Rev. Lett. \textbf{91}, 214502 (2003).

\bibitem{xu-quellette-etal:2006}
H. Xu, N. T. Ouellette, and E. Bodenschatz,
Phys. Rev. Lett. \textbf{96} 114503 (2006).

\bibitem{friedrich:2003}
R.~Friedrich,
Phys. Rev. Lett. \textbf{90},  084501 (2003).

\bibitem{corrsin:1959}
S. Corrsin,
Advances in Geophysics, Vol. 6, ed. F.N. 
Freinkel and P.A. Sheppard (New York Academic), pp 161 (1959).

% \bibitem{belinicher-lvov:1987} 
% V. I. Belinicher, V.S. L'vov,
% JETP \textbf{66}, 309 (1987).

\bibitem{gotoh-fukayama-etal:2002}
T. Gotoh, D. Fukayama and T. Nakano,
Phys. Fluids \textbf{14}, 1065 (2002)

\bibitem{benzi-ciliberto-etal:1993}
R. Benzi, S. Ciliberto, R. Tripiccione, C. Baudet, F. Massaioli and S. Succi,
Phys. Rev. E 48, \textbf{R29} (1993).

\bibitem{ott-mann:2005}
S. Ott, J. Mann,
New J. Phys. {\bf 7}, 142 (2005).


\end{thebibliography}
\end{document}